\documentclass[twocolumn,tighten,times]{aastex63}
\usepackage{xcolor}
\usepackage{apjfonts}

\newcommand{\chjwl}{$ch_{\rm JWL}$}
\newcommand{\nhjwl}{$nh_{\rm JWL}$}
\newcommand{\cnjwl}{$cn_{\rm JWL}$}

\newcommand{\nfehhk}{$n$(MP):$n$(MR)}

\newcommand{\nrgb}{$n$(FG):$n$(SG)}
\newcommand{\nrgbca}{$n$(Ca-w):$n$(Ca-s)}
\newcommand{\pchjwl}{$\parallel$$ch_{\rm JWL}$}
\newcommand{\pnhjwl}{$\parallel$$nh_{\rm JWL}$}

\newcommand{\str}{Str\"omgren}

\newcommand{\vvhbmag}{$-$2.5 mag $\leq$ $V - V_{\rm HB}$ $\leq$ 3.0 mag}
\newcommand{\cnwave}{$\lambda$3883}
\newcommand{\chwave}{$\lambda$4250}
\newcommand{\nhwave}{$\lambda$3360}

\newcommand{\hkjwl}{$hk_{\rm JWL}$}

\newcommand{\fehhk}{[Fe/H]$_{hk}$}
\newcommand{\cfech}{[C/Fe]$_{ch}$}
\newcommand{\nfenh}{[N/Fe]$_{nh}$}
\newcommand{\cfe}{[C/Fe]}
\newcommand{\nfe}{[N/Fe]}
\newcommand{\feh}{[Fe/H]}

\newcommand{\ebv}{$E(B-V)$}

\newcommand{\cfemv}{$d\mathrm{[C/Fe]}/M_V$}
\newcommand{\ciso}{$^{12}$C/$^{13}$C}

\shorttitle{Carbon Depletion}
\shortauthors{Lee}

\begin{document}

\title{Carbon Abundance of Globular Cluster M22 (NGC 6656) and the Surface Carbon Depletion Rates of the Milky Way Globular Clusters}

\author[0000-0002-2122-3030]{Jae-Woo Lee}
\affiliation{Department of Physics and Astronomy, Sejong University, 209 Neungdong-ro, Gwangjin-Gu, Seoul 05006, Republic of Korea,
jaewoolee@sejong.ac.kr, jaewoolee@sejong.edu}

\begin{abstract}
It is well known that metal-poor red giant branch (RGB) stars show variations in some elemental abundances, including carbon, due to the internal mixing accompanied by their own in situ CN cycle in the hydrogen burning shell.  With our new photometric carbon abundance measurements of RGB stars in M22 and other globular clusters (GCs) in our previous studies, M5, M3, and M92, we derive the carbon depletion rates against the $V$ magnitude, \cfemv, for individual populations in each GC. We find the metallicity dependence of the carbon depletion rates, \cfemv\ $\propto$ $-$0.25[Fe/H]. Our results also suggest that the carbon depletion rates of the second generation (SG) of stars are larger than those of the first generation (FG) of stars in our sample GCs, most likely due to different internal temperature profiles with different initial helium abundances between the FG and SG. Our results can provide critical constraints both on understanding the mixing efficiency in the theoretical models, which is largely unknown, and on interpretation of the observational carbon abundance evolution of the bright halo RGB stars.
\end{abstract}

\keywords{Stellar populations (1622); Population II stars (1284); Hertzsprung Russell diagram (725); Globular star clusters (656); Chemical abundances (224); Stellar evolution (1599); Red giant branch (1368)}

\section{INTRODUCTION}
In a generally accepted globular cluster (GC) formation scenario, the second generation (SG) of the stars formed out of interstellar media polluted by the first generation (FG) of the stars \citep[e.g.,][]{dercole08}. The SG stars in normal GCs exhibit different elemental abundance patterns than the FG stars do \citep[][and references therein]{cassisi20, gratton19}. For example, the nitrogen enhancement and carbon depletion in the main sequence (MS) and the lower red giant branch (RGB) stars of the typical GC SGs  could be explained as a natural consequence of the CN cycle occurred in the previous generation of stars. Photometrically, variations in the carbon and nitrogen abundances leave distinctive hallmarks through strong absorption band features of NH, CN, CH molecules, which help to lucidly define multiple stellar populations (MSPs) of GCs \citep{lee17, lee19, lee21a, milone17}.

\begin{figure*}
\epsscale{1.}
\figurenum{1}
\plotone{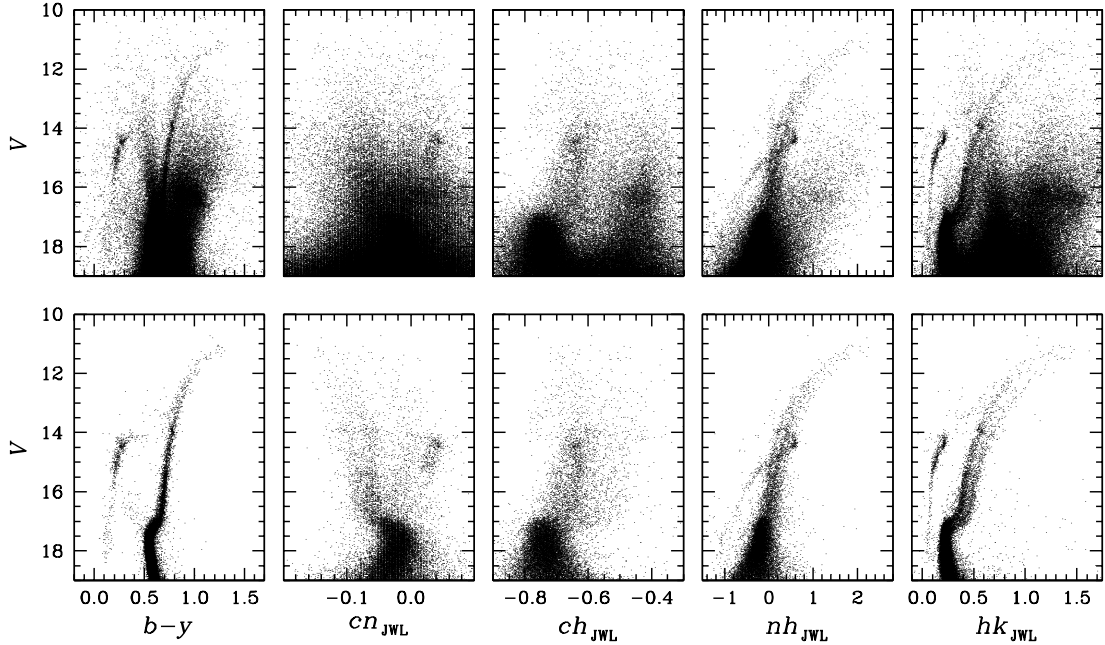}
\caption{(Top) CMDs of M22 in our science field.
(Bottom) CMDs for M22 proper-motion member stars.}\label{fig:cmd}
\end{figure*}

However, the interpretation of carbon and nitrogen abundances of the bright RGB stars can be somewhat complicated due to internal mixing episodes accompanied by their own in situ CN cycle in the hydrogen shell burning region, which significantly alters their initial carbon and nitrogen abundances. Many candidates for this non-canonical extra mixing have been suggested and the thermohaline mixing is considered to be the most promising one since it appears to reproduce observational constraints, although the detailed mixing efficiency used in the theoretical models appears to be largely unknown \citep[e.g.,][]{charbonnel07, lee10, angelou11}.

A few elements, such as $^7$Li, $^{12}$C, \ciso, and $^{14}$N, have been frequently used to fine-tune the unknown mixing parameters. It is believed that neither the \nfe\ nor \ciso\ are proper probes to investigate the internal mixing, since the \nfe\ of the initially nitrogen enhanced stars (i.e., the SG stars) may not be significantly affected by any extra mixing \citep{angelou11}, and the \ciso\ ratio can rapidly attain near-equilibrium value and quickly saturated by only moderate amount of mixing \citep[e.g.,][]{sneden86}. The reliable lithium abundance can be obtained with the \ion{Li}{1} resonance doublet at 6707.78 \AA\ in a rather clean spectral region. However, lithium may not be ideal element to investigate the mixing. Due to its fragility, lithium can be heavily destructed at the RGB bump (RGBB) luminosity level and does not provide any useful information for bright GC RGB stars \citep[e.g., see Figure~1 of][and references therein]{angelou15}. For the same reason, the degree of lithium depletion is known to be extremely sensitive to the stellar models \citep{lattanzio15}.

The last element standing, carbon, is abundant enough that it cannot be completely exhausted via the CN cycle in GC RGB stars. Unfortunately, there is no observable atomic transitions in the optical passband and one should rely on the diatomic molecular absorption bands, such as CH and CN, to derive \cfe. A great deal of effort has been directed to determine carbon abundances by others \citep[e.g.,][]{briley04, smith03, sneden86}.

During the past decade, we developed a new photometric system combined with the robust and self-consistent theoretical fine model grids with various parameter sets to simultaneously measure the key elements of GC MSPs, \feh, \cfe, and \nfe, even in the very crowded field, where traditional spectroscopic observations cannot be applied \citep{lee17, lee19, lee21b, lee22, lee23, lee21a}.

In this Letter, we present a photometric \feh, \cfe, and \nfe\ study for the metal-complex GC M22 \citep{lee09, lee15, lee16, lee20, marino09, marino11}. With our new \cfe\ measurements of M22 and those of our previous studies for M3, M5, and M92, we will discuss the surface carbon depletion rates of the Milky Way GCs to shed more light on the quantitative details of the internal mixing processes.

\section{OBSERVATIONS AND DATA REDUCTION}\label{s:reduction}
The journals of observations for M22 are given in \cite{lee15, lee20}.
In 2019, we obtained additional photometric data for our JWL34 and \str\ $y$, $b$ for M22, with the total integration times of 6200, 260, and 650 s, respectively, in three nights from June 27 to July 8 using the KPNO 0.9 m telescope equipped with a 2 $\times$ 2k CCD chip, providing a field of view (FOV) of 21\arcmin $\times$ 21\arcmin.

The raw data handling was described in detail in our previous works \citep{lee15, lee17}. The photometry of M22 and standard stars were analyzed using DAOPHOTII, DAOGROW, ALLSTAR and ALLFRAME, and COLLECT-CCDAVE-NEWTRIAL packages \citep{pbs87, pbs94}. Finally, we derived the astrometric solutions for individual stars using the Gaia Early Data Release 3 \citep[EDR3;][]{gaiaedr3} and the IRAF IMCOORS package.

In order to select M22 member stars, we made use of the proper-motion study from the Gaia EDR3. We derived the mean proper-motion values using iterative sigma-clipping calculations, finding that, in units of milliarcsecond per year, ($\mu_{\rm RA}\times\cos\delta$, $\mu_{\rm decl.}$) = (9.792, $-$5.611) with standard deviations along the major axis of the ellipse of 0.756 mas yr$^{-1}$ and along the minor axis of 0.679 mas yr$^{-1}$. We emphasize that our mean proper motion for M22 is in good agreement with previous value derived by the Gaia Collaboration \citep[e.g.,][]{gaiahelmi}. We considered stars within 3$\sigma$ from the mean values to be M22 proper-motion member stars.

\begin{figure*}
\epsscale{1.}
\figurenum{2}
\plotone{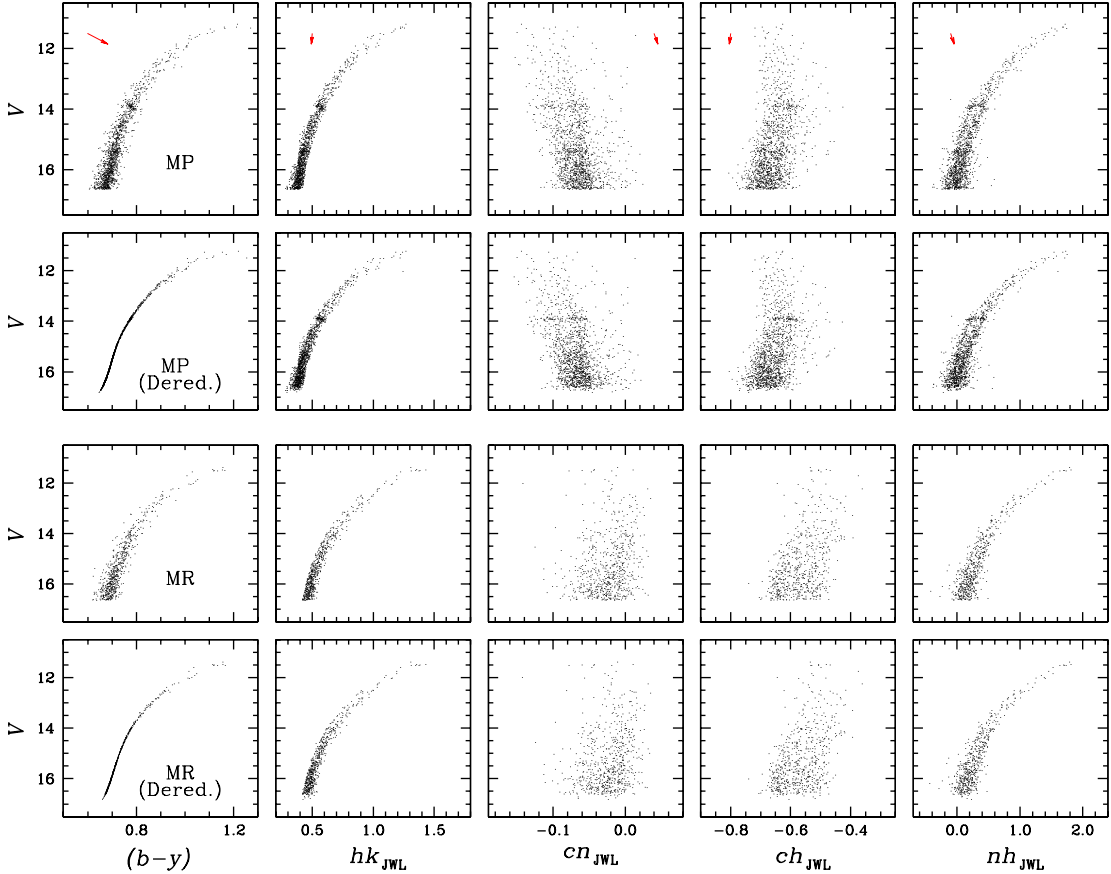}
\caption{CMDs of member RGB stars with \vvhbmag\ in M22. In the top panels, we show differential reddening vectors of individual color indices for $\Delta$\ebv\ = 0.1 mag with red arrows.
}\label{fig:dered}
\end{figure*}

\section{Photometric Indices and Color--Magnitude Diagrams}
Throughout this work, we will use our own photometric indices \citep[see also][]{lee19, lee21a, lee22}, defined as
\begin{eqnarray}
{{hk}}_{\mathrm{JWL}} &=& ({\mathrm{Ca}}_{\mathrm{JWL}}-b)-(b-y),\\
{{cn}}_{\mathrm{JWL}} &=& {JWL}39-{\mathrm{Ca}}_{\mathrm{JWL}},\\
{{ch}}_{\mathrm{JWL}} &=& ({JWL}43-b)-(b-y),\\
{{nh}}_{\mathrm{JWL}} &=& ({JWL}34-b)-(b-y).
\end{eqnarray}
The \hkjwl\ index is a good photometric measure of metallicity, while the \nhjwl, \cnjwl, and \chjwl\ indices are measures of NH absorption band at \nhwave, CN at \cnwave, and CH at \chwave\ \AA, respectively \citep[e.g., see][]{lee09, lee15, lee22, lee23, lee21a}.

In Figure~\ref{fig:cmd}, we show our color--magnitude diagrams (CMDs) of M22 using our color indices. The figure shows that our procedure of selecting cluster's member stars by using the Gaia proper motion works excellently and the most of the foreground and background off-cluster field stars are removed. As we showed in our previous works \citep{lee09, lee15, lee20}, a bimodal \hkjwl\ distribution of M22 RGB stars can be clearly seen. Also M22 has very broad \cnjwl, \chjwl, and \nhjwl\ RGB sequences due to the variations in the carbon and nitrogen abundances as will be discussed below \citep[see also][]{hesser79, lee15}.

We select the M22 member RGB stars of our interest with \vvhbmag\ in order to derive their \fehhk, \cfech, and \nfenh. In Figure~\ref{fig:dered}, we show CMDs of the metal-poor (MP) and the metal-rich (MR) RGB stars in M22. The classification of the two populations will be discussed below \citep[see also][]{lee09, lee15}. At a glance, the differences in the \cnjwl, and \chjwl\ CMDs between the MP and MR populations are noticeable, suggesting that they have different carbon and nitrogen abundances.

The mean interstellar reddening value toward M22 is rather large, \ebv\ = 0.32 \citep[][2010 version]{harris96}, and our M22 CMDs may be vulnerable to differential reddening effect across our science field. We attempt to correct the potential differential reddening effect in the following way. First, we calculated the extinction law for our filter system. We calculated the color excess of individual color indices using our filter transmission functions, synthetic spectra of RGB stars for \feh\ = $-$1.70 dex with a primordial CNO and helium abundances as described below, and the interstellar extinction law by \citet{mathis90}. Through our calculations, we obtained $E(b-y)$ = 0.820\ebv, $E$(\cnjwl) = 0.053\ebv, $E$(\chjwl) = $-$0.043\ebv, $E$(\nhjwl) = 0.539\ebv, and $E$(\hkjwl) = $-$0.067\ebv. In Figure~\ref{fig:dered}, we show differential reddening vectors of individual color indices for $\Delta$\ebv\ = 0.1 mag. Except for the $(b-y)$, the reddening vectors do not point to broaden our observed color indices (i.e., the reddening vectors align along the RGB sequences). Therefore, it is thought that differential reddening across M22 does not significantly affect our results \citep[see also][]{lee15}. In order to examine our claim, we derived the mean $(b-y)$ fiducial sequences of each population and calculated $\Delta$\ebv\ values of individual RGB stars from their $E(b-y)$. Finally, we calculated the reddening corrected color indices and we show them in Figure~\ref{fig:dered}. Except for the $(b-y)$ color, our solutions hardly reduce the RGB widths, suggesting that the broad RGB widths in \hkjwl, \cnjwl, \chjwl, and \nhjwl\ are mainly due to variations in the metallicity, carbon, and nitrogen abundances.

\begin{figure}
\epsscale{1.1}
\figurenum{3}
\plotone{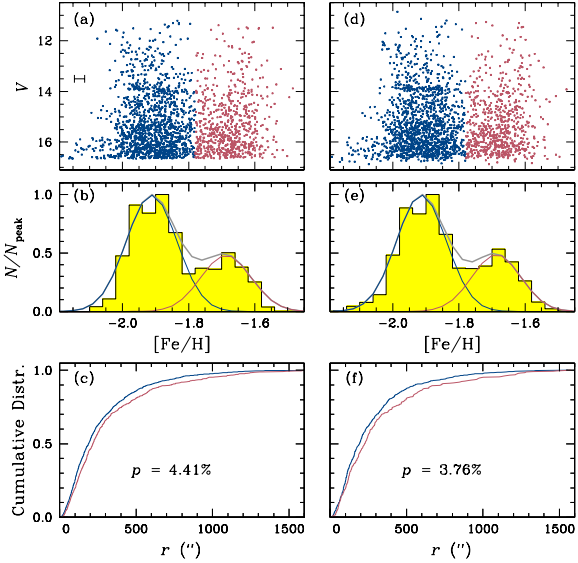}
\caption{(a) The \fehhk\ distribution of M22 RGB stars. The blue and red colors denote the MP and MR populations.
(b) The histogram of the \fehhk\ along with results returned from our EM estimator.
(c) CRDs of each RGB population.
(d--f) Same as (a-c) but for the \fehhk\ from the dereddened \hkjwl\ colors.
}\label{fig:feh}
\end{figure}

\begin{figure*}
\epsscale{1.}
\figurenum{4}
\plotone{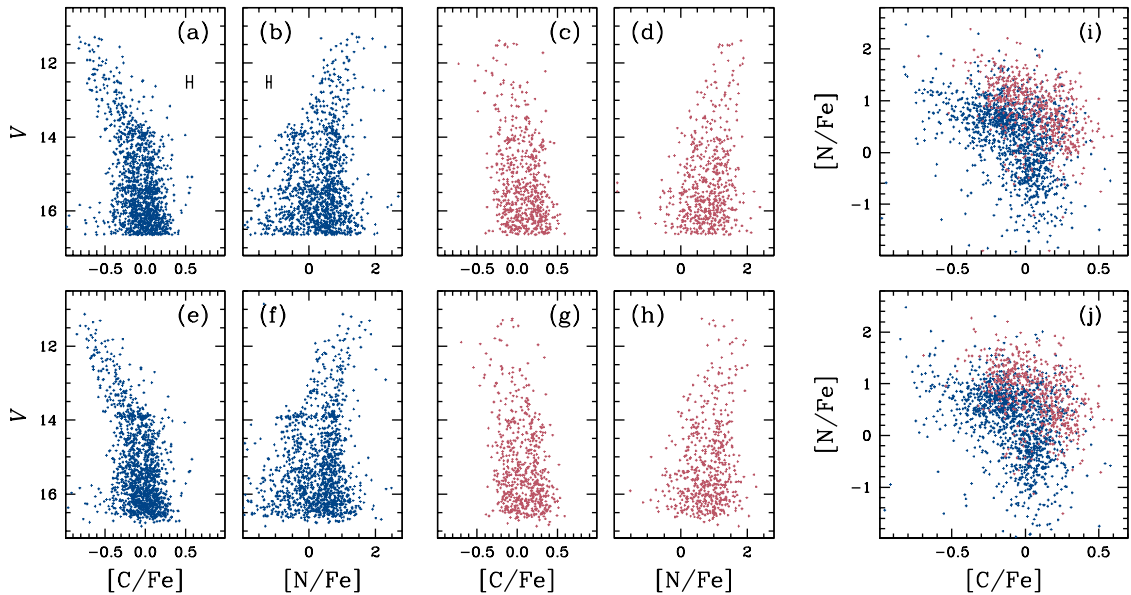}
\caption{(Top) Plots of \cfech\ and \nfenh\ of individual M22 RGB stars against $V$ magnitude. The blue and red colors denote the MP and MR populations. The error bars in (a) and (b) are for typical measurement uncertainties in our \cfech\ and \nfenh\ derivation, respectively.
(Bottom) Same as the top panels but the results from the dereddened color indices.}\label{fig:cfenfe}
\end{figure*}

\section{Metallicity, Carbon, and Nitrogen Abundances}
We derive metallicity of individual RGB stars in M22 using our \hkjwl\ measurements. We retrieved the model isochrones for [Fe/H] = $-$2.1, $-$1.9, $-$1.7, and $-$1.5 dex, $Y$ = 0.247(0.248), 0.275, and 0.300 with [$\alpha$/Fe] = +0.4 dex, and the age of 12.5 Gyr from a Bag of Stellar Tracks and Isochrones \citep{basti21}. We adopted different CNO abundances, [C/Fe] = ($-$0.6, $\Delta$[C/Fe] = 0.2 ,0.6), [N/Fe] = ($-$0.8, $\Delta$[N/Fe] = 0.4 , 1.6), and [O/Fe] = (0.1, 0.3, 0.5) for each model grid. Note that our presumed CNO abundances do not affect our photometric metallicity \citep{lee22}. We constructed 97 model atmospheres and synthetic spectra for each chemical composition from the lower main sequences to the tip of RGB sequences using ATLAS12 \citep{kurucz11} and the latest version of MOOGSCAT \citep{moogscat, sneden74}. As we discussed in our previous works \citep{lee21a, lee23}, the latest version of MOOGSCAT \citep{moogscat} takes proper care of Rayleigh scattering from neutral hydrogen (RSNH) atoms with a nonlocal thermodynamic equilibrium treatment of the source function, which is important to calculate continuum opacities for our short-wavelength indices such as our JWL34 (i.e., \nhjwl), due to a $\lambda^{-4}$ dependency of the RSNH cross section \citep[see][and references therein]{lee21a}.

The photometric metallicity of individual RGB stars can be calculated using the following relation \citep[also see Appendices of][]{lee21a}
\begin{eqnarray}
{\rm [Fe/H]}_{hk} &\approx& f_1({{hk}}_{\mathrm{JWL}},~ M_V).\label{eq:FeH}
\end{eqnarray}
We obtained the mean \fehhk\ = $-$1.839 $\pm$ 0.003 dex ($\sigma$ = 0.129) and we show our results in Figure~\ref{fig:feh}. We emphasize that our photometric metallicity does not show any gradient against $V$ magnitude and exhibits a bimodal distribution as already well known \citep{lee09, lee15, lee16, marino09, marino11}.

In order to perform populational tagging for metallicity, we applied an expectation-maximization (EM) algorithm for a two-component Gaussian mixture model on our \fehhk\ distribution. Stars with $P$(\fehhk$|x_i) \geq$ 0.5 from the EM estimator correspond to the MP population, where $x_i$ denotes the individual RGB stars, while those with $P$(\fehhk$|x_i)$ $<$ 0.5  correspond to the MR population.\footnote{Note that the MP of current study is corresponding to the Ca-w \citep{lee15} and the G1 \citep{lee20}, while the MR to the Ca-s and the G2, respectively.} We obtained the populational number ratio of \nfehhk\ = 68.4:31.6 ($\pm$1.4) and our current result is consistent with our previous result, \nrgbca\ = 70:30 \citep{lee15}. We show the metallicity distributions of the two populations in Figure~\ref{fig:feh}(b). We obtained \fehhk\ = $-$1.914 $\pm$ 0.002 ($\sigma$ = 0.070) for the MP population and $-$1.676 $\pm$ 0.002 ($\sigma$ = 0.050) for the MR population. The difference between the two metallicity groups, $\Delta$\fehhk\ = 0.238 $\pm$ 0.003 dex, is in excellent agreement with those in spectroscopic analyses \citep[e.g.,][]{lee16}. We also examined the cumulative radial distributions (CRDs) and we show them in Figure~\ref{fig:feh}(c). We performed Kolmogorov--Smirnov tests and obtained that the MP and MR populations are most likely drawn from the different parent CRDs \citep[see also][]{lee09, lee15}.

With our differential reddening estimates as mentioned above, we calculate the metallicity from the dereddened \hkjwl\ colors, finding \fehhk\ = $-$1.840 $\pm$ 0.003 dex ($\sigma$ = 0.129), $-$1.915 $\pm$ 0.002 ($\sigma$ = 0.069), $-$1.676 $\pm$ 0.002 ($\sigma$ = 0.058) for the mean value, the MP, and MR populations with the identical populational number ratio \nfehhk\ = 68.4:31.6 ($\pm$1.4). As we pointed out above, we emphasize again that the differential reddening does not appear to affect our results.

Using the following relations \citep{lee21b, lee21a, lee23},
\begin{eqnarray}
{\rm [C/Fe]}_{ch} &\approx& f_2({{ch}}_{\mathrm{JWL}},~{\mathrm{[Fe/H]}_{hk}}, ~ M_V), \\
{\rm [N/Fe]}_{nh} &\approx& f_3({{nh}}_{\mathrm{JWL}},~{\mathrm{[Fe/H]}_{hk}}, ~ M_V),
\end{eqnarray}
we derive the photometric \cfech\ and \nfenh\ of each population in M22. In Figure~\ref{fig:cfenfe}, we show \cfech\ and \nfenh\ against the $V$ magnitude with and without the differential reddening correction, suggesting that the differential reddening correction does not significantly affect our photometric \cfech\ and \nfenh. Our results clearly show that the carbon abundance decreases and nitrogen abundance increases in RGB stars brighter than the RGBB ($V$ $\approx$ 14.0 mag for M22; \citealt{lee15}) as can be seen in our previous studies of other Galactic GCs, M5, M3, and M92 \citep{lee21a, lee21b, lee23}.

In Figures~\ref{fig:cfenfe}(i--j), the distributions of the MP and MR on the \cfech\ versus \nfenh\ plane are different in the sense that at a given carbon abundance the nitrogen abundance of the MP RGB stars are lower, confirming previous results of \citet{marino11} and \citet{lee15}. As \citet{marino11} noted, without any discernible oxygen abundance differences, the difference in the total CNO abundances between the two populations in M22 is likely responsible for this separation.

In Figure~\ref{fig:cfenfe}, we emphasize that the degrees of the carbon depletion between the M22 MP and MR RGB stars brighter than the RGBB appears to be quite different. On the other hand, the degree of the nitrogen enhancement is not as clear as that of the carbon depletion, at least due to the dependence of the nitrogen enhancement on their initial nitrogen abundances \citep[e.g.,][]{angelou11}.

We performed subpopulational tagging for the MP and MR populations.
Due to the carbon depletion and the nitrogen enhancement in RGB stars brighter than the RGBB, neither \cfech\ nor \nfenh\ may be the proper probes for the populational tagging. Instead, we perform populational tagging using our \pchjwl\ and \pnhjwl, which are defined as
\begin{eqnarray}
\parallel ch_{\rm JWL} &\equiv& \frac{ch_{\rm JWL} - ch_{\rm JWL,red}}
{ch_{\rm JWL,red}-ch_{\rm JWL,blue}},\\
\parallel nh_{\rm JWL} &\equiv& \frac{nh_{\rm JWL} - nh_{\rm JWL,red}}
{nh_{\rm JWL,red}-nh_{\rm JWL,blue}},
\end{eqnarray}
where the subscripts denote the fiducials of the red and the blue sequences of individual color indices \citep[see also][]{lee21a}. With these procedures, the curvatures of the RGB sequence on the \chjwl\ and \nhjwl\ CMDs can be effectively removed. Then we can perform populational tagging for the bright RGB stars on these normalized color indices \citep[e.g.,][]{lee20}.
We made histograms for the (\pchjwl\ $-$ \pnhjwl), which is roughly equivalent to [C/N], showing at least two distinctive peaks in each histogram. We applied EM algorithms for each population and obtained subpopulational number ratios of \nrgb\ = 49.0:51.0 ($\pm$2.4) for the MP and 50.0:50.0 ($\pm$3.5) for the MR.

\begin{figure*}
\epsscale{1.}
\figurenum{5}
\plotone{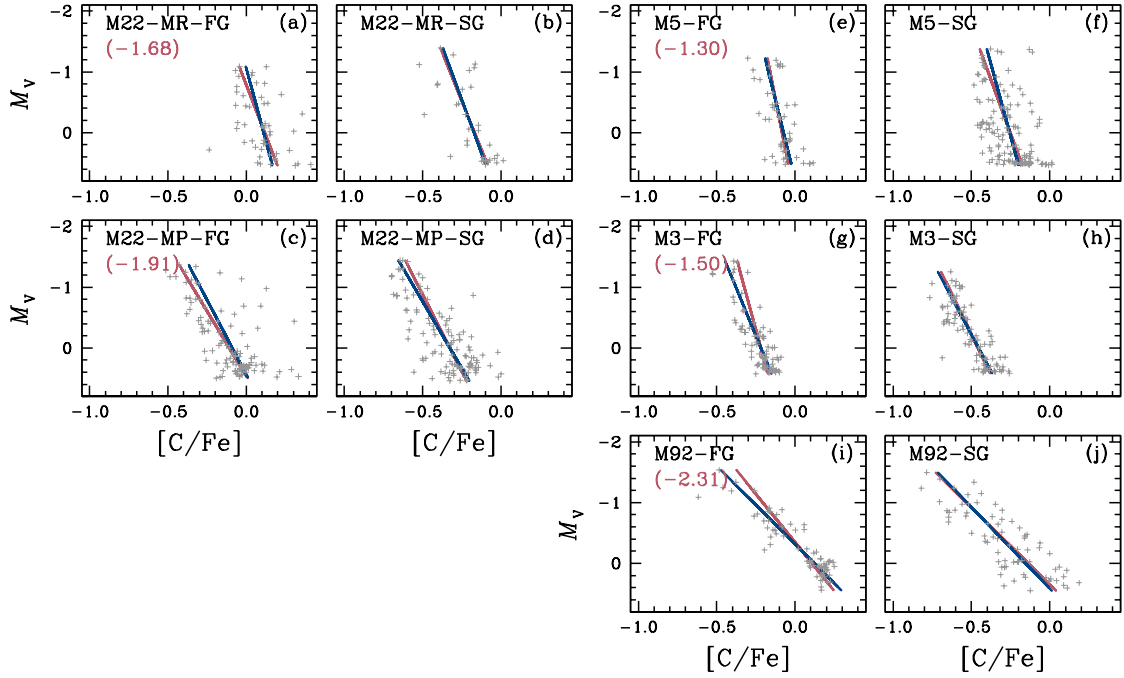}
\caption{Plots of \cfech\ vs. $M_V$ of individual populations of M22, M5, M3, and M92 for RGB stars brighter than their RGBB $V$ magnitudes. The blue and red solid lines denote the least-squares fitting and the robust fitting by minimizing absolute deviation, respectively. The numbers in parentheses are the mean metallicity of each GC.}\label{fig:gcs}
\end{figure*}

With our subpopulations, we derived $V$ magnitude dependence of carbon abundance depletion, \cfemv. In order to derive the slopes of such relation, we used a ordinary least-squares fit and a robust fit that minimizes absolute deviation \citep[MEDFIT:][]{nure}. We show our results in Figure~\ref{fig:gcs} and Table~\ref{tab:slope}. As shown, the M22 MP population has steeper gradients than the MR does, mainly due to difference in the mean metallicity between the two populations \citep{charbonnel07}. Also importantly, the SG subpopulations appear to have slightly steeper gradients than the FG, although the differences are not statistically significant.

\begin{deluxetable*}{lcccccc}[t]
\tablenum{1}
\tablecaption{Magnitude Dependence of Carbon Abundance.\label{tab:slope}}
\tablewidth{0pc}
\tablehead{
\multicolumn{1}{c}{} &
\multicolumn{1}{c}{Pop.} &
\multicolumn{1}{c}{[Fe/H]\tablenotemark{a}} &
\multicolumn{1}{c}{\fehhk} &
\multicolumn{1}{c}{$d$\cfe/$dM_V$\tablenotemark{b}} &
\multicolumn{1}{c}{$d$\cfe/$dM_V$\tablenotemark{c}} &
\multicolumn{1}{c}{$d$\cfe/$dM_V$\tablenotemark{d}}
}
\startdata
M22-MR & FG   &         & $-$1.68 &  0.104 $\pm$ 0.033 &  0.150 $\pm$ 0.088 & \\
       & SG   &         &         &  0.142 $\pm$ 0.029 &  0.156 $\pm$ 0.074 & \\
M22-MP & FG   &         & $-$1.91 &  0.203 $\pm$ 0.024 &  0.235 $\pm$ 0.090 & \\
       & SG   &         &         &  0.227 $\pm$ 0.020 &  0.198 $\pm$ 0.092 & \\
\hline
M5     & FG   & $-$1.29 & $-$1.30 &  0.096 $\pm$ 0.019 &  0.077 $\pm$ 0.056 & \\
       & SG   &         &         &  0.106 $\pm$ 0.017 &  0.141 $\pm$ 0.089 & \\
M3     & FG   & $-$1.50 & $-$1.50 &  0.159 $\pm$ 0.012 &  0.105 $\pm$ 0.049 & 0.236 $\pm$ 0.033 \\
       & SG   &         &         &  0.206 $\pm$ 0.016 &  0.187 $\pm$ 0.055 & \\
M92    & FG   & $-$2.31 & $-$2.31 &  0.391 $\pm$ 0.022 &  0.315 $\pm$ 0.066 & 0.227 $\pm$ 0.045\\
       & SG   &         &         &  0.374 $\pm$ 0.032 &  0.396 $\pm$ 0.119 & \\
\enddata
\tablenotetext{a}{\citet[][2010 version]{harris96}
\tablenotetext{b}{A least-squares fitting.}
\tablenotetext{c}{A fitting by minimizing absolute deviation.}
\tablenotetext{d}{\citet{smith03}.}
}
\end{deluxetable*}

\begin{figure*}
\epsscale{1.1}
\figurenum{6}
\plotone{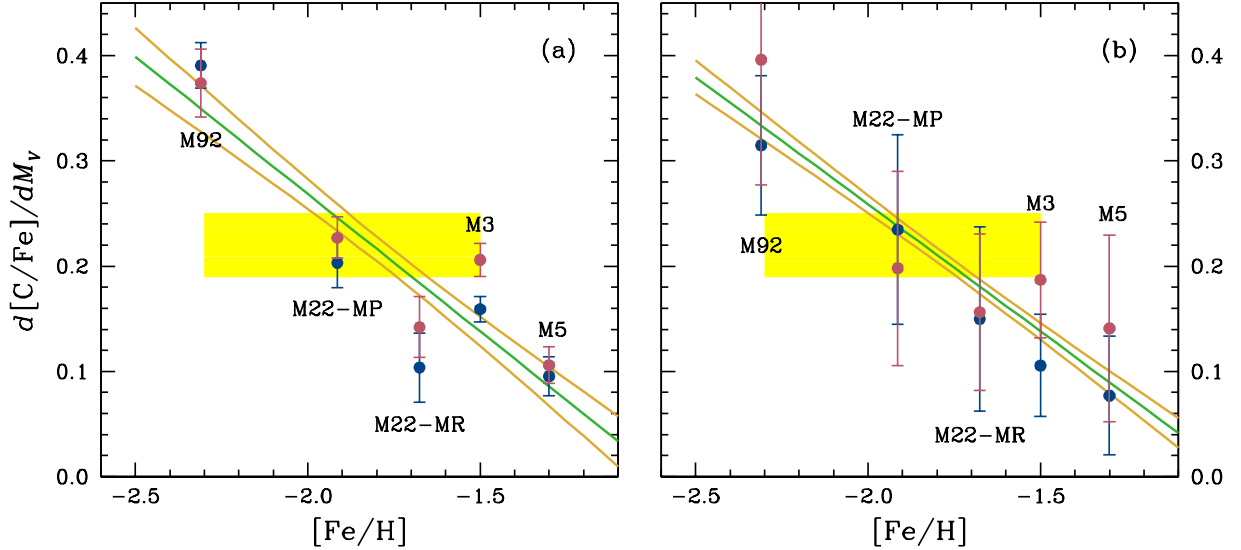}
\caption{(a) A plot of \cfemv\ vs. \feh\ returned from the least-squares fitting (blue solid lines in Figure~\ref{fig:gcs}). The blue and red colors denote the FG and SG, respectively. The green and gold solid lines show a linear regression with confidence intervals of 95\%. The yellow shade box indicates the mean value by \citet{smith03}.
(b) Same as (a) but for results returned from the robust fitting by minimizing absolute deviation (red solid lines in Figure~\ref{fig:gcs}).}\label{fig:grad}
\end{figure*}

\section{Surface Carbon Depletion Rates of Milky Way Globular Clusters}
It is well known that the mixing efficiency depends on metallicity since the hydrogen burning shell will be hotter and wider with decreasing metallicity \citep[e.g.,][]{church14}. We investigate the degree of the carbon depletion against $V$ magnitude, \cfemv, with expand samples of our previous studies, M5 \citep[\feh\ = $-$1.29;][]{lee21b}, M3 \citep[$-$1.50;][]{lee21a}, and M92 \citep[$-$2.31;][]{lee23}. As with M22, we derive the slopes of the relations using two different methods. We show the results in Figure~\ref{fig:gcs} and Table~\ref{tab:slope}. In the table, the \cfemv\ of M3 and M92 by \citet{smith03} are also listed. \citet{smith03} did not distinguish the FG and SG in their analyses and their slopes for M3 and M92 are slightly different from our values.

In Figure~\ref{fig:grad}, we show plots of \cfemv\ versus [Fe/H] for the case of a least-squares fit (Case 1; column (5) of Table~\ref{tab:slope}) and a robust fit, which minimizes absolute deviation (Case 2; column (6) of Table~\ref{tab:slope}), respectively. We obtained following relations
\begin{eqnarray}
 \frac{d\mathrm{[C/Fe]}}{dM_V} &=& -0.261(\pm0.043)\mathrm{[Fe/H]} - 0.253(\pm0.077),\\
 \frac{d\mathrm{[C/Fe]}}{dM_V} &=& -0.241(\pm0.037)\mathrm{[Fe/H]} - 0.224(\pm0.066),
\end{eqnarray}
and, not surprisingly, our results clearly show that the \cfemv\ depends on metallicity. Note that \citet{smith03} obtained the mean value of \cfemv\ $\approx$ 0.22 $\pm$ 0.03 from three GCs (M92, NGC 6397, and M3) and the MP halo giants. On the other hand, \citet{martell08} argued that the carbon depletion rate\footnote{We note that the carbon depletion rate by \citet{martell08}, $\Delta$[C/Fe]/$\Delta$t, is different from our value, \cfemv.} becomes doubled from \feh\ = $-$1.3 to $-$2.3.

As we mentioned above, the slopes of the \cfemv\ for the FG and SG appear to be slightly different for M5, M3, and M92, in the sense that the SG tends to have a steeper slope. We argue that different initial helium abundances between the FG and SG are responsible for the difference in the slopes. Due to the high helium abundances of the SG stars, which can be inferred from their RGBB $V$ magnitudes \citep[e.g.,][]{lee20, lee23, lee21a}, the SG RGB stars have high temperature at the hydrogen burning shell, which causes slightly faster carbon destruction through the CN cycle \citep[e.g.,][]{church14}.

\section{SUMMARY}
We investigated the photometric metallicity, carbon, and nitrogen abundances of the metal-complex globular cluster M22. Our results confirmed previous results that M22 contains at least two MSPs with heterogeneous metallicities \citep{lee09, lee15, lee16, marino09, marino11}.

We obtained \fehhk\ = $-$1.839 $\pm$ 0.003 dex ($\sigma$ = 0.129), $-$1.914 $\pm$ 0.002 ($\sigma$ = 0.070), and $-$1.676 $\pm$ 0.002 ($\sigma$ = 0.050) for the all RGB, the MP, and MR RGB populations, respectively. Applying differential reddening correction does not affect our photometric metallicity measurements, no larger than 0.001 dex in the mean values.

Our \cfech, \nfenh\ measurements of individual populations in M22 showed that each population showed evidences of the CN process accompanied by deep mixing episodes with different degrees.

With our carbon abundance measurements for M22 and other GCs in our previous studies, M5 \citep{lee21b}, M3 \citep{lee21a}, M92 \citep{lee23}, we investigated the surface carbon depletion rates, \cfemv, for Milky Way GCs against metallicity, finding \cfemv\ $\propto$ $-$0.25[Fe/H]. We also argued that the carbon depletion rates of the SG are larger than those of the FG, most likely due to different initial helium abundances between the FG and SG, which cause different internal temperature profiles.

Our results presented here can provide critical constraints both on understanding the mixing efficiency in the theoretical models, which is largely unknown, and on interpretation of the carbon abundances evolution seen from the bright halo RGB stars.

\acknowledgements
{J.-W.L.\ acknowledges financial support from the Basic Science Research Program (grant No.\ 2019R1A2C2086290) through the National Research Foundation of Korea (NRF) and from the faculty research fund of Sejong University in 2022. He also thanks the anonymous referee for encouraging comments.}

\facilities{SMARTS: 1.0 m (STA), WIYN: 0.9 m (HDI, S2KB), Gaia.}

\end{document}